\documentclass[aps,prl,twocolumn,superscriptaddress]{revtex4-2}
\usepackage{amsmath} 
\usepackage{amssymb}
\usepackage{bm}
\usepackage{dcolumn}
\usepackage{graphicx}
\graphicspath{{.}{./figs/}} 
\usepackage{xcolor}
\usepackage{changes} 

\begin{document}

\title{Tolerance and breakdown of topological protection in a disordered waveguide}

\author{Kiyanoush Goudarzi}
\email{kgoudar@ncsu.edu}
\affiliation{Department of Electrical Engineering, Pohang University of Science and Technology (POSTECH), 37673 Pohang, Korea}
\affiliation{Department of Electrical and Computer Engineering, North Carolina State University, Raleigh, NC 27695, USA}
\author{Moonjoo Lee}
\email{moonjoo.lee@postech.ac.kr}
\affiliation{Department of Electrical Engineering, Pohang University of Science and Technology (POSTECH), 37673 Pohang, Korea}

\date{\today}

\begin{abstract}
We consider a disordered waveguide consisting of trivial dielectric and non-trivial magnetically anisotropic material. 
A topologically-protected edge mode appears owing to the broken time-reversal symmetry of the non-trivial lattice. 
While the edge mode maintains under other position and radius disorders, the protection is immediately broken by applying a radius disorder to the non-trivial lattice.
This breakdown originates from donor and acceptor modes occupying the topological bandgap.
Furthermore, via the calculation of the Bott index, we show that Anderson localization occurs as a metal conducting gap changes to a topological gap along with increasing disorders.
\end{abstract}

\maketitle


Topologically protected state is generally known to be stable and robust against perturbations.
This topological protection (TP) holds significant importance in various disciplines, including photonics~\cite{Ozawa2019, liu2022topological}, condensed matter~\cite{ghorashi2023topological, li2022imaging}, atomic physics~\cite{cooper2019topological, Walter2023}, and acoustics~\cite{yang2015topological}. 
In particular, TP provides a solution for certain problems in photonics through the unique transport property of the topological edge mode. 
For instance, in a nanophotonic device, the transmission of an electromagnetic (EM) wave is often affected by backreflection and scattering losses. 
The topological waveguide enables the transmission of reflection-free waves, even in the presence of substantial structural disorder~\cite{Wang2008reflection, Wang2009observation}.  
Besides, photonic systems with TP have been excellent testbeds for exploring non-Hermitian physics~\cite{weidemann2022topological, liang2022dynamic}, non-linear optics~\cite{jung2022thermal, jurgensen2023quantized}, higher-order band topology~\cite{xie2021higher}, and Floquet physics~\cite{rechtsman2013photonic, nagulu2022chip}.

This TP, however, can be influenced or even broken by specific environmental conditions. 
In condensed matter physics, extensive studies were performed regarding the breakdown of the TP.
For example, the breakdown was observed in graphene edges due to antidots and upstream modes in an electrostatic regime~\cite{marguerite2019imaging, moreau2021upstream}. 
The TP breakdown also occurred in a two-dimensional electron gas by enhanced vacuum fluctuations~\cite{appugliese2022breakdown}, and a smooth edge potential broke the TP via edge reconstruction~\cite{wang2017spontaneous}. 
In contrast, such investigations in photonics are relatively scarce. 
One study indicated TP breaking by transitioning from the amorphous to crystalline structural state of the constitutive material Ge$_{2}$Sb$_{2}$Te$_{2}$~\cite{cao2019dynamically}. 
Another study revealed a TP breakdown using rotational symmetry in the unit cell of periodic Kekul\'{e} patterns~\cite{Tian2023}.

Here, we explore the tolerance and breakdown of TP of an edge mode in a disordered waveguide.
Our topological waveguide (TW) consists of a trivial lattice interfaced with a non-trivial lattice.
When both the position and radius disorders are applied to the trivial lattice, the unidirectionality of the edge mode sustains, exhibiting the TP nature of the mode. 
The unidirectionality also maintains under the position disorder of the non-trivial lattice.
However, when the radius disorder is applied to the non-trivial lattice, all different localized defect modes prevent the formation of the topological bandgap, which breaks the edge mode. 
Moreover, as the radius disorder increases, a frequency domain of bulk modes becomes a non-trivial bandgap, accompanying Anderson localization (AL) of EM waves in the topological regime. 
Such topological nature is characterized with the Bott index (BI), a topological invariant obtained from the real-space wavefunctions.

\begin{figure}[t]
\includegraphics[width=3.3in]{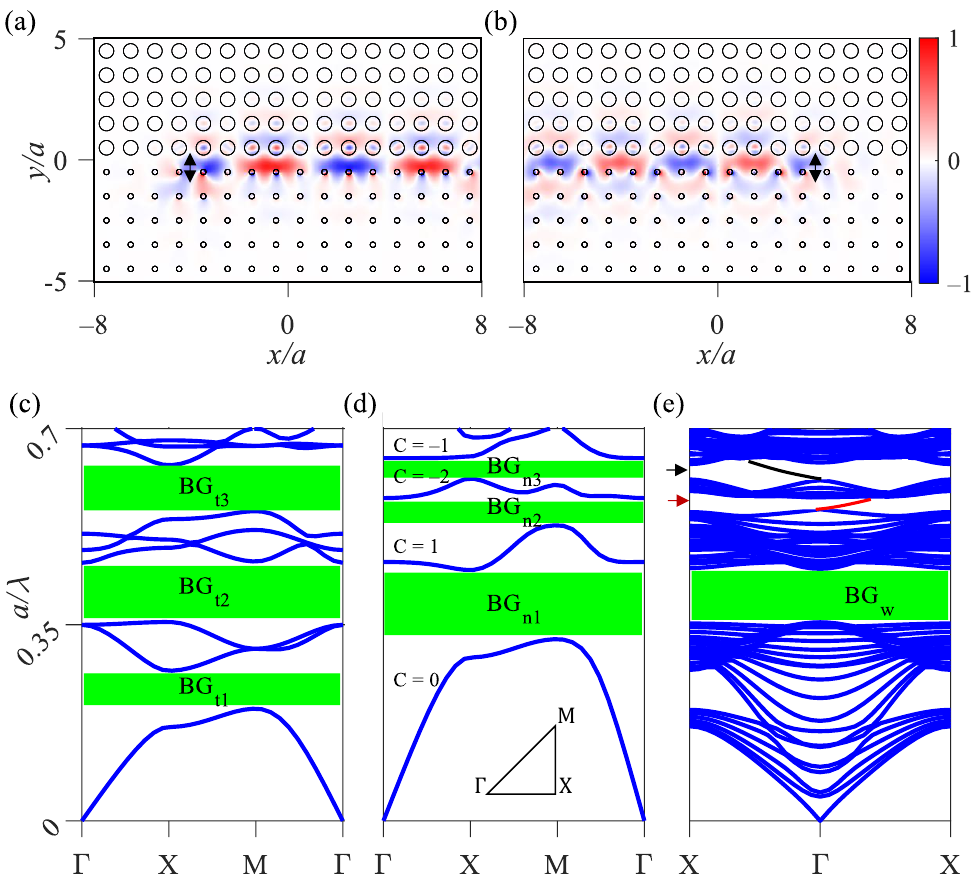}
	\caption{
		Normalized $E_{z}(x, y)$ at $a/\lambda = 0.570$ in (a) and $0.630$ in (b). 
		Black arrows denote the location of light source.
		(c)--(d) Dispersion diagrams of trivial and non-trivial lattices.
		Green regions show photonic bandgaps and $\Gamma$, M, and X represent vortices of the irreducible Brillouin zone. 
		Lattice constant, wavelength, and Chern number are referred to as $a$, $\lambda$, and C, respectively. 
		(e) Dispersion diagram of TW. 
		Red and black curves show edge modes with positive and negative group velocities. Red and black arrows indicate $a/\lambda = 0.570$ and $0.630$.
	}
	\label{fig:fig1} 
\end{figure}

Our TW includes a trivial and a non-trivial mirror. 
The trivial mirror contains dielectric rods with a relative permittivity of $\epsilon_\textrm{t} = 16$, a relative permeability of $\mu_\textrm{t} = 1$, and a radius of $R_\textrm{t} = 0.30a$ in air where $a$ is the lattice constant. 
The trivial lattice corresponds to the upper half of Figs.~\ref{fig:fig1}(a) and (b).
The non-trivial lattice consists of the magneto-optical material of yttrium-iron-garnet (YIG) ferrite rods with a relative permittivity of $\epsilon_\textrm{n} = 15$ and a permeability of $\mu_\textrm{n} = \bm{\mu}$ with a radius of $R_\textrm{n} = 0.11a$ in air (lower half of Figs.~\ref{fig:fig1}(a) and (b)). 
Near an operating frequency of 4.5~GHz, the anisotropic permeability $\bm\mu$ under applying a static magnetic field is

\begin{equation}\label{eq:1}
	\bm{\mu} =
		\begin{bmatrix}
			\mu & i\kappa & 0 \\
  			-i\kappa & \mu & 0 \\
  			0 & 0 & \mu_0
		\end{bmatrix},
\end{equation}

\noindent 
where $\mu = 14\mu_0$ and $\kappa = 12.4\mu_0$, where $\mu_0$ is the vacuum permeability~\cite{Wang2008reflection}.

Fig.~\ref{fig:fig1}(c) shows the dispersion diagram of the trivial mirror with the three trivial TM bandgaps of BG$_\textrm{t1}$, BG$_\textrm{t2}$, and BG$_\textrm{t3}$. 
The first two gaps of BG$_\textrm{t1}$ and BG$_\textrm{t2}$ are Mie bandgaps with TM$_{01}$ and TM$_{11}$ modes, respectively. 
The TM$_{01}$ and TM$_{11}$ modes constitute pure Mie bandgaps that exhibit high tolerance to the position and radius disorderings, and thus the penetration of EM waves into the trivial mirror is well suppressed. 
Differently from BG$_\textrm{t1}$ and BG$_\textrm{t2}$, both Mie and Bragg scatterings generate BG$_\textrm{t3}$ with the TM$_{21}$ mode, less tolerant to the disorderings, causing that the EM waves would penetrate gently more into the trivial lattice than BG$_\textrm{t1}$ and BG$_\textrm{t2}$~\cite{goudarzi2022super}.

In the non-trivial mirror, we find three bandgaps of BG$_{\textrm{n1}}$, BG$_{\textrm{n2}}$, and BG$_\textrm{n3}$ in the dispersion diagram of Fig.~\ref{fig:fig1}(d). 
The lowest bandgap of BG$_\textrm{n1}$ is a trivial Mie bandgap with TM$_{01}$ mode.  
The second and third bandgaps are non-trivial topological gaps.
Due to the broken time-reversal symmetry in the non-trivial lattice, the high-symmetry points of M and $\Gamma$ in the irreducible Brillouin zone break, resulting in the creation of BG$_\textrm{n2}$ and BG$_\textrm{n3}$, respectively~\cite{Wang2008reflection, goudarzi2022calculation}.
Breaking the time-reversal symmetry originates from the fact that the anisotropic material of YIG has imaginary off-diagonal elements in the permeability tensor with $\bm\mu \neq \bm\mu^{T}$, where $T$ is the transpose operation~\cite{caloz2018electromagnetic}. 
The calculations of the Chern number and dispersion diagram are elaborated in Ref.~\cite{supplemental}.

We proceed to the calculation of Chern numbers in both the trivial and non-trivial lattices.
The Chern numbers over the first Brillouin zone are zero for all bands of the trivial lattice. 
The zero value of the Chern numbers is based on preserving the time-reversal symmetries~\cite{Wang2008reflection, Wang2020universal, goudarzi2022calculation}. 
As contrary, we obtain the Chern numbers of $0, 1, -2,$ and $-1$ for the first four bands, from low to high frequency, in the non-trivial lattice~\cite{supplemental}.

Each band's Chern number constitutes the value of the Chern number for each bandgap, i.e., the bandgap's Chern number (C$_\textrm{g}$) is defined as the sum of bands' Chern number below the bandgap~\cite{Wang2008reflection, Wang2020universal, goudarzi2022calculation}.
As a consequence, we obtain C$_{\textrm{g}} = 0, 1,$ and $-1$ for BG$_{\textrm{n1}}$, BG$_{\textrm{n2}}$, and BG$_{\textrm{n3}}$, respectively. 
The overlap between BG$_\textrm{t2}$ and BG$_\textrm{n1}$ gives rise to the creation of the bandgap of BG$_\textrm{w}$ at normalized frequencies $0.353<a/\lambda<0.451$ in the dispersion diagram of the TW: This results in no edge mode, because the difference of C$_\textrm{g}$ of BG$_\textrm{n1}$ and that of BG$_\textrm{t2}$ is zero. 
Differently from BG$_\textrm{w}$, the overlap between BG$_\textrm{t3}$ and the two bandgaps of BG$_\textrm{n2}$ and BG$_\textrm{n3}$ creates two unidirectional edge modes at $0.556 < a/\lambda < 0.578$ and $0.613 < a/\lambda < 0.637$ with positive and negative group velocities, respectively, as shown in Fig.~\ref{fig:fig1}(e). 
The edge modes are visualized in Figs.~\ref{fig:fig1}(a) and (b). 
The out-of-plane electric fields, $E_{z}(x, y)$, at $a/\lambda = 0.570$ and $0.630$ represent the propagation of EM waves to the right and left directions.

\begin{figure*}[t]
\includegraphics[width=6.3in]{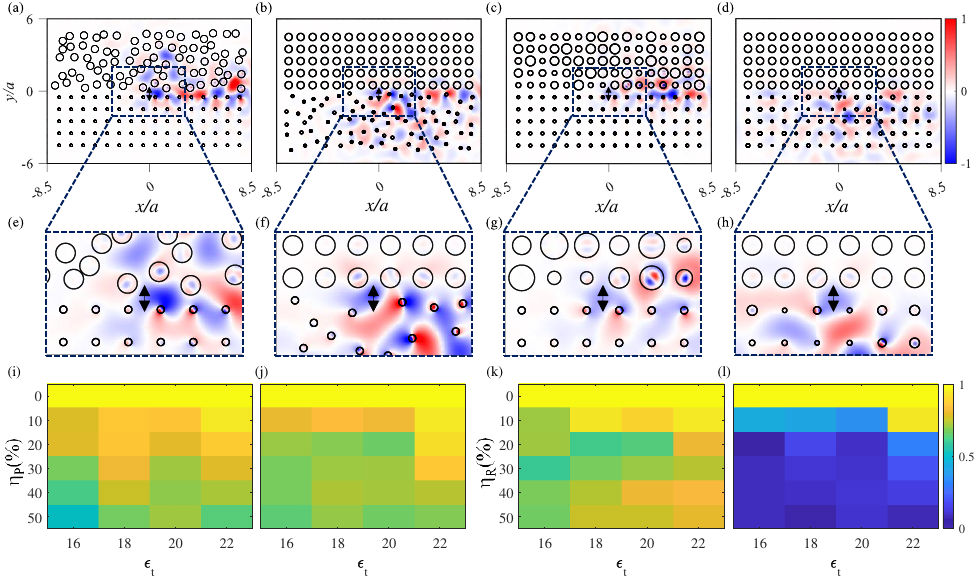}
	\caption{
	\label{fig:fig2} 
	(a)--(h) Normalized $E_{z}(x, y)$ under position and radius disorders.
	(a), (e) Trivial mirror with $\eta_\textrm{P} = 40\%$, (b), (f) non-trivial mirror with $\eta_\textrm{P} = 40\%$, (c), (g) trivial mirror with $\eta_\textrm{R} = 40\%$, and (d), (h) non-trivial mirror with $\eta_\textrm{R} = 40\%$.
	TM polarized dipole sources are denoted as black arrows at the interface of the mirrors with a normalized frequency $a/\lambda = 0.570$.
	Transmission $T_\textrm{LR}$ when position disorder is applied to (i) trivial and (j) non-trivial lattices, and radius disorder to (k) trivial and (l) non-trivial lattices for several $\epsilon_\textrm{t}$.
	$a/\lambda = 0.570$ for $\epsilon_{\textrm{t}}=16, 18$, and $20$ and $a/\lambda=0.550$ for $\epsilon_{\textrm{t}}=22$.}
\end{figure*}

Next, we explore the impact of disorders to the unidirectionality of the edge modes. 
We consider two types of disorderings, in the position and radius of the rods. 
The disordered position of the rod $i$ is defined as $(x^{i}, y^{i})$, where $x^{i} = x_{0}^{i} + \sigma_\textrm{P} F_{x}^{i}$ and $y^{i} = y_{0}^{i} + \sigma_\textrm{P} F_{y}^{i}$ with the original position $(x_{0}^{i}, y_{0}^{i})$, the strength of the position disorder is $\sigma_\textrm{P}$, and $F_{x}^{i}$ and $F_{y}^{i}$ are uniformly distributed random variables between $-1$ and $1$ for the $i$th rod along the $x$ and $y$ directions. 
The position disordering parameter is defined as $\eta_\textrm{P} = \sigma_\textrm{P}/a$.
In a similar way, the disordered radius of the rod $i$ in the trivial (non-trivial) lattice is $R_\textrm{t(n)}^{i} = R_\textrm{t(n),0} + \sigma_\textrm{R} F_\textrm{R}^{i}$, where $R_\textrm{t(n),0}$, $\sigma_\textrm{R}$, and $F_\textrm{R}^{i}$ stand for the original radius of the rod in the trivial (non-trivial) lattice, strength of the radius disordering, and the uniformly distributed random parameter over the interval of $[-1, 1]$. 
We define the parameter of the radius disordering as $\eta_{\textrm{R}} = \sigma_\textrm{R}/R_{\textrm{t(n)}, 0}$.

We first obtain $E_{z}(x, y)$ under the influence of the position disordering. 
The parameters of the trivial mirror are $R_\textrm{t,0} = 0.30a$, $\epsilon_\textrm{t} = 18$, and $\mu_{\textrm{t}}=1$ and those of the non-trivial mirror are $R_\textrm{n,0} = 0.11a$, $\epsilon_\textrm{n} = 15$, and $\mu_{\textrm{n}} = \bm{\mu}$. 
The normalized frequency of the TM polarized dipole source is $a/\lambda = 0.570$ where the edge mode of a positive group velocity appears. 
The calculation results are shown when the disorder is $\eta_\textrm{P} = 40$\% in either the trivial or non-trivial mirror (Figs. \ref{fig:fig2}(a), (b), (e), and (f)): The unidirectionality of the EM waves is preserved at this substantial strength of disorder.

We explore the unidirectionality more quantitatively by calculating the transmittance $T_\textrm{LR}$ of the edge mode. 
We judge that the EM wave has the unidirectionality transmission when $T_{\textrm{LR}} > 0.5$. 
Figs.~\ref{fig:fig2}(i) and (j) show $T_{\textrm{LR}}$ at several disorderings and $\epsilon_\textrm{t}$. 
Each transmittance is obtained by averaging the results of 100 numerical simulations. 
Under applying the position disorders to both the trivial and non-trivial mirrors, the unidirectionality is preserved and $T_\textrm{LR}$ moderately decreases as the disorder increases from $\eta_\textrm{P}=0$ to $50\%$.

Similar behavior is observed when the radius disorder is applied to the trivial lattice (Figs.~\ref{fig:fig2}(c), (g), and (k)). 
However, the result is entirely different when the radius of non-trivial lattice is disordered.
Figs.~\ref{fig:fig2}(d) and (h) show that the EM waves propagate to $+x, -x$, and $-y$ directions and the penetration to the trivial lattice ($+y$ direction) is suppressed.
$T_\textrm{LR}$ abruptly decreases from 1 to 0.4 at $\eta_{\textrm{R}}=10\%$, and to $\sim0$ at $\eta_{\textrm{R}}=20\%$---the unidirectionality is broken completely (Fig.~\ref{fig:fig2}(l)).

\begin{figure*}[t]
\includegraphics[width=6.3in]{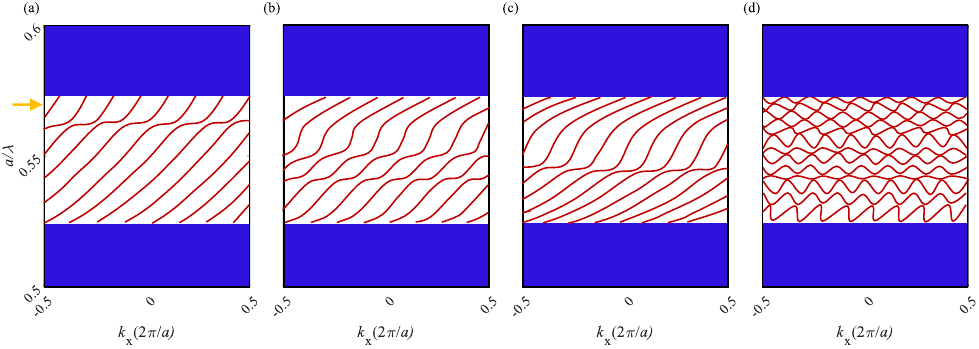}
	\caption{
	Dispersion diagrams of TW with $5 \times 8$ dielectric rods and $5 \times 8$ YIG rods.  
	(a) $\eta_\textrm{P} = 40\%$ to trivial lattice, (b) $\eta_\textrm{P} = 40\%$ to non-trivial lattice, (c) $\eta_\textrm{R} = 40\%$ to trivial lattice, and (d) $\eta_\textrm{R} = 40\%$ to non-trivial lattice. 
	Blue regions are bulk modes. 
	Yellow arrow indicates $a/\lambda = 0.570$.
	}
	\label{fig:fig3} 
\end{figure*}

We describe the origin of the tolereance and breakdown of the edge mode.
Fig.~\ref{fig:fig3} shows the dispersion diagrams of the TW, containing a supercell with $5 \times 8$ of dielectric rods with $\epsilon_\textrm{t} = 18$, $\epsilon_\textrm{n}=15$, and $5 \times 8$ YIG rods of $\bm{\mu}$~\cite{supplemental}.  
When the position disorder $\eta_{\textrm{P}}=40\%$ is applied to either the trivial or non-trivial lattice (Figs.~\ref{fig:fig3}(a) and (b)), the edge mode with a positive group velocity still exists and thus the unidirectionality of the EM wave maintains. 
This point is associated with the mode profile in the rods and the interference of the EM waves. 
In Figs.~\ref{fig:fig2}(e) and (f), we identify that the localized modes inside each dielectric and YIG rod under the position disordering are TM$_{21}$ and TM$_{11}$, respectively.
The unidirectionality to the $+x$ direction is a result of preserving the localized mode inside each rod in the non-trivial lattice. 
The gentle penetration of EM waves to the $+y$ (Figs.~\ref{fig:fig2}(a) and (e)) and $-y$ (Figs.~\ref{fig:fig2}(b) and (f)) directions is owing to the random increase in the distances between some rods. 
The increased distances result in decreasing the coupling between quasi-bound states, which makes the EM waves penetrate between the rods~\cite{nojima2013quantitative}.
However, all identical TM$_{11}$ modes in YIG rods sustain the slope of the edge mode.

In case of radius disorders, defect modes affect the directionality of the edge mode. 
In this configuration, each disordered rod corresponds to a point defect of the waveguide.
The point defects with larger (smaller) radii, increase (decrease) the effective refractive index, resulting in the appearance of donor (acceptor) modes. 
The donor (acceptor) modes, defect modes, fall into the adjacent bandgap from the upper edge (lower edge) of the gap, resulting in filling the bandgap with the defect modes. 
In other words, donor and acceptor modes experience red and blue shifts, respectively; the acceptor modes between BG$_\textrm{t2}$ and BG$_\textrm{t3}$ (radius disorder to trivial lattice), acceptor modes between BG$_\textrm{n1}$ and BG$_\textrm{n2}$ (radius disorder to non-trivial lattice), and donor modes between BG$_\textrm{n2}$ and BG$_\textrm{n3}$ (radius disorder to non-trivial lattice) generate the bulk modes that occupy the gap.

\begin{figure}[!t]
\includegraphics[width=3.3in]{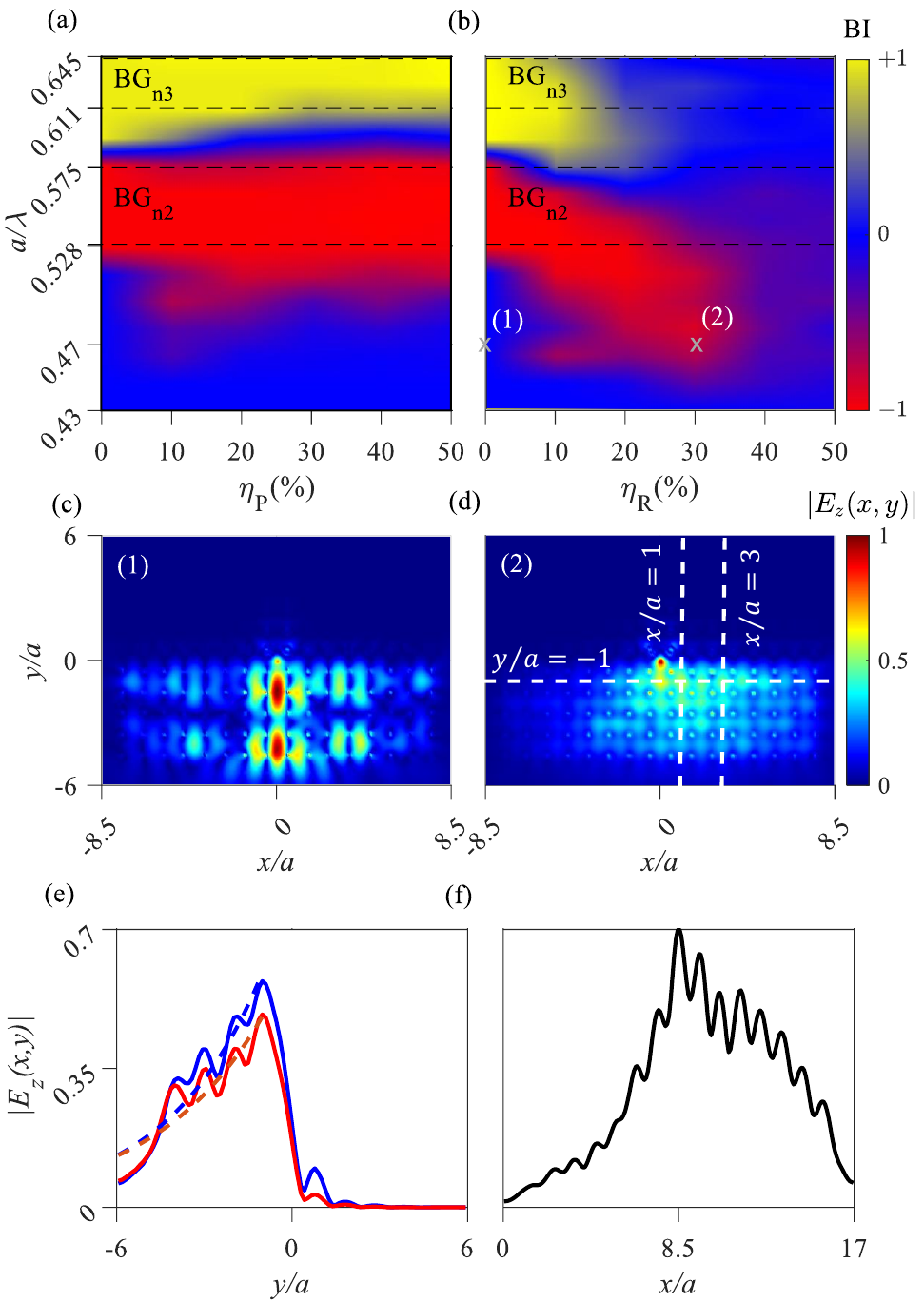}
	\caption{
	Bott index of $6 \times 6$ YIG rods over a frequency range of $0.430\leq a/ \lambda \leq 0.660$ under (a) position and (b) radius disorders.
	(c) $|E_{z}(x, y)|$ at the location (1) of bulk modes. 
	(d) $|E_{z}(x, y)|$ at (2) where TAL happens. 
	Source position of EM wave is the origin.
	We choose $\epsilon_\textrm{t}=24$ so that BG$_\textrm{t3}$ covers $a/\lambda=0.470$.
	(e) $|E_{z}(a, y)|$ (blue solid line) and $|E_{z}(3a, y)|$ (red solid line), and fitting with exponential decay functions (dashed lines).
	(f)  $|E_{z}(x, -a)|$.
	}
	\label{fig:fig4} 
\end{figure}

Note that this description holds for the cases of weak or moderate radius disorders. 
As the disorder increases, the impact of donor mode becomes more dominant than that of accepter mode, because more modes can exist in a rod of larger radius (see the explanation for Fig.~\ref{fig:fig4}(b) below); as the radius disorder increases strongly, the bandgap shifts to red.

The impact of the defect modes is different between occurred in the trivial and non-trivial lattice. 
Given radius disorder in the trivial lattice, the acceptor modes between BG$_\textrm{t2}$ and BG$_\textrm{t3}$ would fill up the gap at $0.556< a/\lambda < 0.578$. 
However, the localized eigenmode in YIG rods does not change, which persists the directionality of the edge mode.
All identical mode profiles in YIG rods are identified in Figs.~\ref{fig:fig2}(c) and (g), and accordingly the positive sign of the group velocity is preserved, as shown in Fig.~\ref{fig:fig3}(c).

In contrast, when the radius disorder is applied to the non-trivial lattice, the topological nature is broken totally. 
As shown in Fig.~\ref{fig:fig3}(d), both the acceptor modes from the frequencies between BG$_\textrm{n1}$ and BG$_\textrm{n2}$ and the donor modes between BG$_{\textrm{n2}}$ and BG$_{\textrm{n3}}$ occupy the gap, and the resulting bulk modes exhibit no directionality. 
The underlying reason of this breakdown is that all the eigenmodes in YIG rods become different under the radius disorder.
All distinct these modes disturb the formation of the topological bandgap of the non-trivial lattice, which herein causes the TP breakdown. 
This feature is also found in Figs.~\ref{fig:fig2}(d) and (h), showing that different modes are localized in every YIG rod.

We further investigate this breakdown of TP with another topological variable, the BI~\cite{liu2020topological,loring2011disordered,bandres2016topological, Agarwala2017topological}.
The BI, obtained from the real-space electric-field distribution, is particularly useful for characterizing disordered structures where the Chern number cannot be defined.
The BI and Chern number manifest the topological nature of a band with nonzero values, having the same absolute value with the opposite sign.
More details of the BI calculation are provided in Ref.~\cite{supplemental}

The calculation results of the BI are presented in Fig.~\ref{fig:fig4}.
We consider a supercell of $6 \times 6$ YIG rods in air under position and radius disorderings over the frequency interval of $0.430\leq a/ \lambda \leq 0.660$. 
The BI at each frequency and disordering is obtained by averaging the results of 50 simulations. 
In case of position disorder, we find two topological bandgaps of BG$_{\textrm{n2}}$ and BG$_{\textrm{n3}}$ with the BI of $-1$ and $+1$, respectively, which do not change as the position disorder increases (Fig.~\ref{fig:fig4}(a)). 
This shows that the edge mode is not influenced by the position disorder of the trivial and non-trivial lattices, agreeing with the results in Figs.~\ref{fig:fig2}(a) and (b), and Figs.~\ref{fig:fig3}(a) and (b).
The localization of TM$_{\textrm{11}}$ mode in all YIG rods gives the same BI, which is independent of position disordering: This supports the unidirectionality of the edge mode.

As shown in Fig.~\ref{fig:fig4}(b), the behavior of the topological bandgap, under the radius disorder of the non-trivial lattice, is very different from that under the position disorder: Both the bandgaps of BG$_\textrm{n2}$ and BG$_\textrm{n3}$ undergo red shift as $\eta_{\textrm{R}}$ increases.
This frequency shift is attributed to the donor modes in the point defects of larger radii. 
Under the radius disorder, defects with both larger and smaller radii are present.
While many bulk modes can appear in the rods of larger radii, the number of modes that can exist in smaller defects would be much less; for instance, certain modes cannot survive if the size of a rod is smaller than the critical radius.
Dominated by the impact of defects with larger radii, the effective index of refraction of the lattice increases, causing the red shift of overall defect modes---the bandgaps shift to red as well. 
More quantitative explanation is offered with the calculations of $E_{z}(x,y)$ and the mode frequencies in Ref.~\cite{supplemental}.

This red shift brings about two phenomena: The breakdown of the TP and the emergence of AL~\cite{schwartz2007transport, li2009topological}.
First, we consider an edge mode at $a/\lambda=0.570$.
As $\eta_{\textrm{R}}$ increases, the topological bandgap is occupied with bulk modes, resulting in the disappearance of the edge mode (BI changes from $-1$ to $0$): This is the reason for the breakdown of the unidirectionality as discussed above.
Second, we concentrate on a region of $0.430<a/\lambda<0.528$ in Fig.~\ref{fig:fig4}(b).
While this domain is occupied with bulk modes at $\eta_{\textrm{R}}=0$, these modes gradually disappear and the topological bandgap emerges as $\eta_{\textrm{R}}$ grows.
Accordingly the edge mode appears in this frequency region, which accompanies not only the unidirectionality of the mode but also the localization of the EM waves under the disorder, which corresponds to AL.
The behavior of AL is revealed in Fig.~\ref{fig:fig4}(e), where $|E_{z}(a, y)|$ and $|E_{z}(3a, y)|$ are fitted with the exponential decay function of $E_{0} \cdot \exp{(-|y|/\xi)}$.
The localization length is given by $\xi$.
From the fitting, we obtain $\xi \simeq 3.3a$ for $|E_{z}(a, y)|$ and $\xi \simeq 3.8a$ for $|E_{z}(3a, y)|$, and this exponentially decaying feature proves the genuine nature of AL. 
Summarizing the phenomena here, as the disorder increases, transition from the metal conducting gap to non-trivial bandgap happens, in tandem with a disorder-induced emergence of AL in this topological regime.

We finally remark two points regarding our work. 
First, as far as we are aware, our work is for the first time to explore the impact of radius disorder in a topological waveguide. 
While previous studies were done with position disorders~\cite{mansha2017robust, xiao2017photonic, yang2019topological}, we study how the radius disorders affect the edge mode, making it possible to manifest the TP breakdown.
Second, we herein envision the possibility of a similar study in the optical and infrared (IR) frequency domain.
The deployed YIG material exhibits the nature of broken time-reversal frequency near 4.5~GHz~\cite{Wang2008reflection, Wang2009observation}.
In order to investigate such topological behavior in the optical domain, one can utilize a honeycomb lattice including a dielectric-helical waveguide, in an ambient medium with a refractive index of $1.45${~\cite{rechtsman2013photonic}. 
The symmetry breaks in the helical direction, which acts like time-reversal symmetry breaking in YIG, leading to the unidirectional EM wave around the lattice at optical frequencies. 
In the IR region, one can make use of dielectric rods with a refractive index of $3.42$ in a honeycomb pattern in air, revealing pseudo-time reversal symmetry preservation. 
This would result in the generation of unidirectional chiral edge states in the IR domain~\cite{Wu2015}.

In conclusion, we have studied the tolerance and breakdown of TP in a disordered waveguide.   
Both the position and radius disorders are applied to either the trivial or non-trivial lattice.
The edge mode disappears under the radius disorder of non-trivial lattice, because all different donor and acceptor modes prevent the creation of the topological gap.  
Moreover, through the calculation of a topological variable of the BI, we show that EM waves are localized in a certain frequency region, which is AL effect associated with the topological gap.  
Our work offers new understanding on the one-way light propagation in nanophotonics, and also gives an insight to the development of topological optical circuits and integrated photonic devices.

\begin{acknowledgments}
We thank H.~G.~Maragheh for helpful discussions. 
We acknowledge the support from BK21 FOUR program and Educational Institute for Intelligent Information Integration, National Research Foundation (Grant No.~2019R1A5A1027055), Samsung Electronics Co., Ltd (IO201211-08121-01), and Samsung Science and Technology Foundation (SRFC-TC2103-01).
\end{acknowledgments}

Our data are available at https://doi.org/10.5281/zenodo.8339612.

\bibliography{References}

\clearpage
\twocolumngrid
\begin{center}
\textbf{\large Supplemental Material: Tolerance and breakdown of topological protection in a disordered waveguide}
\end{center}
\setcounter{equation}{0}
\setcounter{figure}{0}
\setcounter{table}{0}
\setcounter{page}{1}
\makeatletter
\renewcommand{\theequation}{S\arabic{equation}}
\renewcommand{\thefigure}{S\arabic{figure}}
\renewcommand{\bibnumfmt}[1]{[S#1]}
\renewcommand{\citenumfont}[1]{S#1}

\section{Dispersion diagram of topological waveguide}

\begin{figure}[!b]
\includegraphics[width=1 \linewidth]{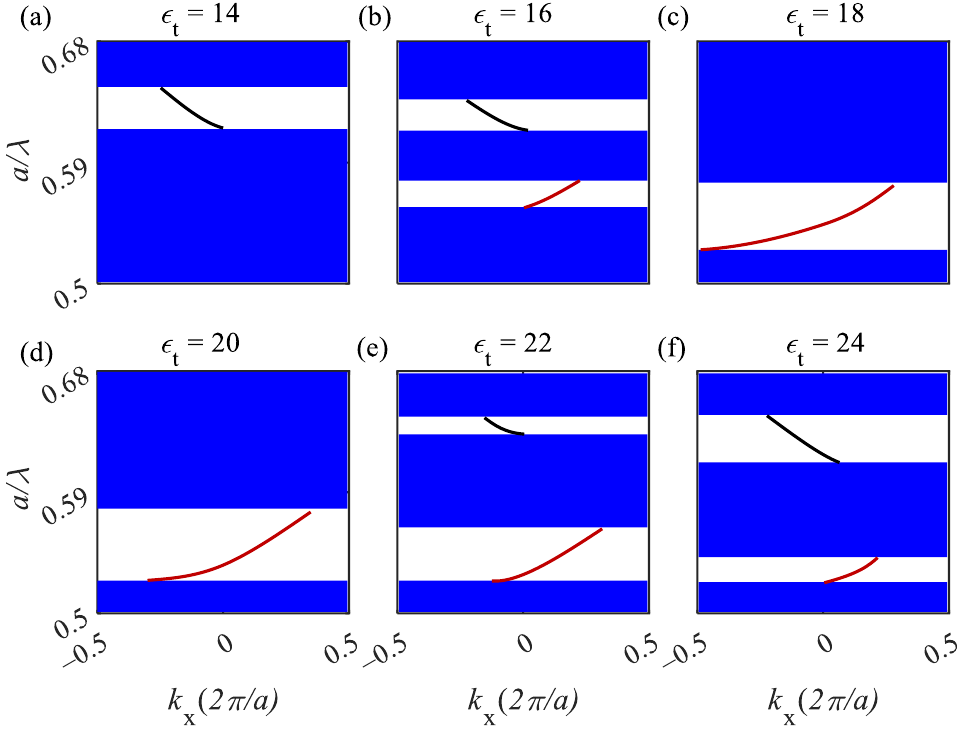}
\caption{\label{fig:S1} 
	The TM band structure of the non-trivial mirror consisting of YIG rods with $R_\textrm{n,0}=0.11a$, $\epsilon_\textrm{n} = 15$, $\mu = 14\mu_0$, and $\kappa = 12.4$ in air and a trivial mirror of dielectric rods in air with $R_\textrm{t,0}=0.3a$ and (a) $\epsilon_\textrm{t}=14$, (b) $\epsilon_\textrm{t}=16$, (c) $\epsilon_\textrm{t}=18$, (d) $\epsilon_\textrm{t}=20$, (e) $\epsilon_\textrm{t}=22$, and (f) $\epsilon_\textrm{t}=24$. 
	The blue color represents bulk modes. 
	Solid black and red curves are edge modes with negative and positive group velocities, respectively.}
\end{figure}

We discuss the transverse magnetic (TM) dispersion diagrams, photonic band structures, of a topological waveguide (TW) containing a trivial mirror with $R_\textrm{t,0}=0.3a$ and $\epsilon_\textrm{t}$ in air and a non-trivial mirror consisting of YIG rods with $R_\textrm{n,0}=0.11a$, $\epsilon_\textrm{n} = 15$, $\mu = 14\mu_0$, and $\kappa = 12.4\mu_0$ in air.
Fig.~\ref{fig:S1} shows the dispersion diagrams of non-trivial bandgaps at $0.5\leq a/\lambda \leq0.68$ with either positive or negative group velocities. 
The positive or negative group velocities over the wavevector $k$\textsubscript{x} represent the unidirectionality of the electromagnetic waves in $+x$ or $-x$ directions, respectively, at the interface between the trivial and non-trivial mirrors. 
Note that the frequency region of the bandgap changes as $\epsilon_{\textrm{t}}$ varies.  
The dispersion diagrams of the TW show an edge mode with negative group velocity and the bandgap's Chern number C\textsubscript{g} = -1 at $\epsilon_\textrm{t} = 14$ (Fig.~\ref{fig:S1}(a)) and an edge mode with positive group velocity and C\textsubscript{g} = +1 at $\epsilon_\textrm{t} = 18$ and 20 (Fig. \ref{fig:S1}(c) and (d)). 
Also, the dispersion diagrams of the TW show two edge modes at $\epsilon_\textrm{t} = 16$, 22, and 24 (Fig.~\ref{fig:S1}(b), (e), and (f)). 
The first mode with higher frequency, negative group velocity and C\textsubscript{g} = -1, and the second mode with lower frequency, positive group velocity and C\textsubscript{g} = +1. 
The blue color in the band structure diagrams stands for bulk modes that are of no unidirectionality.
This calculation is performed using COMSOL Multiphysics software.

\section{The Chern number and Berry curvature calculations}

The Chern number for the band $n$ is calculated in the wavevector space over the first Brillouin zone (BZ). 
BZ is calculated for a unit cell in Fig.~\ref{fig:S2}(a). 
The Chern number is presented as

\begin{equation}\label{eq:1}
	C_n = \frac{1}{2\pi i}\int_{\textrm{BZ}}\textrm{d}^{2}k F^{n}(\mathbf{k}),
\end{equation}

\noindent where

\begin{equation}
	\label{eq:2}
	F^{n}(\mathbf{k}) = \left( \frac{\partial A_y^{nn}}{\partial k_x} - \frac{\partial A_x^{nn}}{\partial k_y} \right),
\end{equation}

\noindent and

\begin{equation}
	\label{eq:3}
	\mathbf{A}^{nn'}(\mathbf{k}) \equiv i \int \textrm{d}^{2}r \, \epsilon_r(\mathbf{r}) \mathbf{E}_{n\mathbf{k}}^{*}(\mathbf{r})\cdot \left[ \nabla_{\mathbf{k}}\mathbf{E}_{n'\mathbf{k}}(\mathbf{r}) \right].
\end{equation}

\begin{figure}[t]
\includegraphics[width=1 \linewidth]{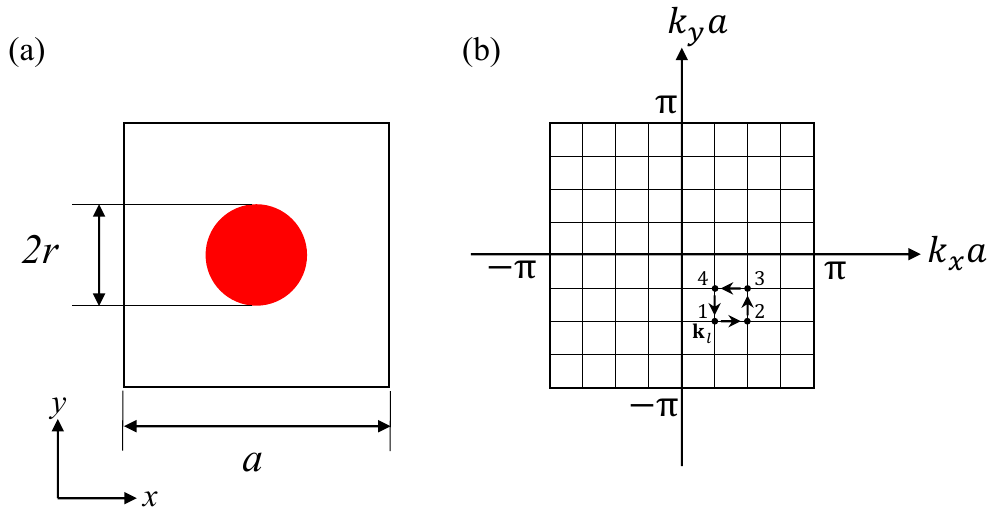}
\caption{
	\label{fig:S2} 
	(a) Unit cell of the periodic structure containing a 2D cylindrical rod in air and (b) the first BZ. 
	$a$ and $r$ are the lattice constant and the radius of the rod, respectively.}
\end{figure}

In the above equations, $F$, $\textbf{A}$, and $\textbf{E}_{n\textbf{k}}$ are the Berry curvature, Berry connection, and electric field (Bloch function), respectively. 
$\epsilon_\textrm{r}(\mathbf{r})$ is the position-dependent dielectric constant and the asterisk stands for the complex conjugation. 
Quantification of the Chern number and Berry curvature requires calculating the eigenfrequencies and eigenvalues over the first BZ bounded at $-\frac{\pi}{a}<k_{x,y}<\frac{\pi}{a}$ as shown in Fig.~\ref{fig:S2}(b), where $a$ is the lattice constant. 
To calculate the Chern number, Eqs.~(\ref{eq:1})--(\ref{eq:3}) are replaced with the following equations

\begin{equation}
	\label{eq:4}
	C_{n}= \frac{1}{2\pi i}\sum_{\mathbf{k}_l\in \textrm{BZ}}F_{xy}^{n}(\mathbf{k}_l)\delta\mathbf{k}_x\delta\mathbf{k}_y.
\end{equation}

\begin{equation}
\begin{aligned}
\label{eq:5}
F_{xy}(\mathbf{k}_l)\delta\mathbf{k}_x\delta\mathbf{k}_y & = (\partial_x A_y (\mathbf{k}_l)-\partial_y A_x (\mathbf{k}_l))\delta\mathbf{k}_x\delta\mathbf{k}_y \\
                                                         &\approx (\Delta_x A_y)\delta\mathbf{k}_y-(\Delta_y A_x)\delta\mathbf{k}_x.
\end{aligned}
\end{equation}

\begin{equation}
\begin{aligned}
\label{eq:6}
A_\lambda(\mathbf{k}_l)\delta\mathbf{k}_\lambda & =  \langle u(\mathbf{k}_l) | \partial_\lambda | u(\mathbf{k}_l) \rangle \delta\mathbf{k}_\lambda\\
                         & = \langle u(\mathbf{k}_l) | u(\mathbf{k}_l+\delta\mathbf{k}_\lambda) -u(\mathbf{k}_l) \rangle \\
                         & =\langle u(\mathbf{k}_l) | u(\mathbf{k}_l+\delta\mathbf{k}_\lambda) \rangle-1.
\end{aligned}
\end{equation}

\noindent where $n$, $A_x$, $A_y$, $|u(\mathbf{k}_l)\rangle$ are the band number (we drop it in the rest of the context), the Berry connection in the $x$ direction, the Berry connection in the $y$ direction, and the Bloch function. 
Utilizing the Taylor series, the Berry connection reads

\begin{equation}
\begin{aligned}
\label{eq:7}
	e^{A_\lambda(\mathbf{k}_l)\partial \mathbf{k}_\lambda} & = e^{\langle u(\mathbf{k}_l) | u(\mathbf{k}_l+\delta\mathbf{k}_\lambda) \rangle-1} \\
	& \approx 1 + \langle u(\mathbf{k}_l) | u(\mathbf{k}_l+\delta\mathbf{k}_\lambda) \rangle-1 \\
	& = \langle u(\mathbf{k}_l) | u(\mathbf{k}_l+\delta\mathbf{k}_\lambda) \rangle,
\end{aligned}
\end{equation}

\begin{equation}
\label{eq:8}
	A_\lambda(\mathbf{k}_l)\partial \mathbf{k}_\lambda = \ln \langle u(\mathbf{k}_l) | u(\mathbf{k}_l+\delta\mathbf{k}_\lambda) \rangle.
\end{equation}

\begin{equation}
\label{eq:9}
	\Delta_\mu A_\lambda(\mathbf{k}_l)\partial \mathbf{k} = A_\lambda(\mathbf{k}_l + \partial \mathbf{k}_\mu)\partial \mathbf{k} - A_\lambda(\mathbf{k}_l)\partial \mathbf{k};   (\mu , \lambda : x, y),
\end{equation}
By substituting Eqs.~(\ref{eq:8}) and (\ref{eq:9}) into Eq.~(\ref{eq:5}), the following equation appears

\begin{equation}
\label{eq:10}
\begin{aligned}
F_{xy}(\mathbf{k}_l)\partial \mathbf{k}_x \partial \mathbf{k}_y = & \ln (\langle u(\mathbf{k}_l
               + \partial \mathbf{k}_x) | u(\mathbf{k}_l
               + \delta\mathbf{k}_x + \delta\mathbf{k}_y)\rangle) \\
               &   - \ln (\langle u(\mathbf{k}_l) | u(\mathbf{k}_l + \delta\mathbf{k}_y)\rangle) \\
               &  - \ln (\langle u(\mathbf{k}_l + \delta\mathbf{k}_y) | u(\mathbf{k}_l + \delta\mathbf{k}_x + \delta\mathbf{k}_y)\rangle)\\
               &  + \ln (\langle u(\mathbf{k}_l) | u(\mathbf{k}_l + \delta\mathbf{k}_x)\rangle).
\end{aligned}
\end{equation}

Because the Chern number is real and integer and the Berry connection is purely imaginary, as a result, the link variables are written as

\begin{equation}
\label{eq:11}
\begin{split}
&U_{12}(\mathbf{k}_l) = \ln\left( \frac{\langle u(\mathbf{k}_l) | u(\mathbf{k}_l + \delta\mathbf{k}_x)\rangle}{|\langle u(\mathbf{k}_l) | u(\mathbf{k}_l + \delta\mathbf{k}_x)\rangle|}\right), \\
&U_{23}(\mathbf{k}_l) = \ln \left( \frac{\langle u(\mathbf{k}_l
               + \partial \mathbf{k}_x) | u(\mathbf{k}_l
               + \delta\mathbf{k}_x + \delta\mathbf{k}_y)\rangle}{|\langle u(\mathbf{k}_l
               + \partial \mathbf{k}_x) | u(\mathbf{k}_l
               + \delta\mathbf{k}_x + \delta\mathbf{k}_y)\rangle|} \right), \\
&U_{34}(\mathbf{k}_l) = \ln \left( \frac{\langle u(\mathbf{k}_l
               + \partial \mathbf{k}_x + \delta\mathbf{k}_y) | u(\mathbf{k}_l
               + \delta\mathbf{k}_y)\rangle}{|\langle u(\mathbf{k}_l
               + \partial \mathbf{k}_x + \delta\mathbf{k}_y) | u(\mathbf{k}_l
               + \delta\mathbf{k}_y)\rangle|} \right),\\
&U_{41}(\mathbf{k}_l) = \ln \left( \frac{\langle u(\mathbf{k}_l
               + \delta\mathbf{k}_y) | u(\mathbf{k}_l)
               \rangle}{|\langle u(\mathbf{k}_l
               + \delta\mathbf{k}_y) | u(\mathbf{k}_l)
               \rangle|} \right),
\end{split}
\end{equation}

Substituting link variables in Eq. (\ref{eq:4}), the Chern number reads

\begin{equation}
	\label{eq:12}
	C_{n}= \frac{1}{2\pi i}\sum_{\mathbf{k}_l\in \textrm{BZ}}\ln (U_{12} U_{23} U_{34} U_{41}).
\end{equation}

As obvious from Eq.~(\ref{eq:12}), to calculate the Chern number at $\textbf{\textrm{k}}_l$, a counterclockwise loop is considered as represented in Fig.~\ref{fig:S2}(b).

\begin{figure}[t]
\includegraphics[width=1 \linewidth]{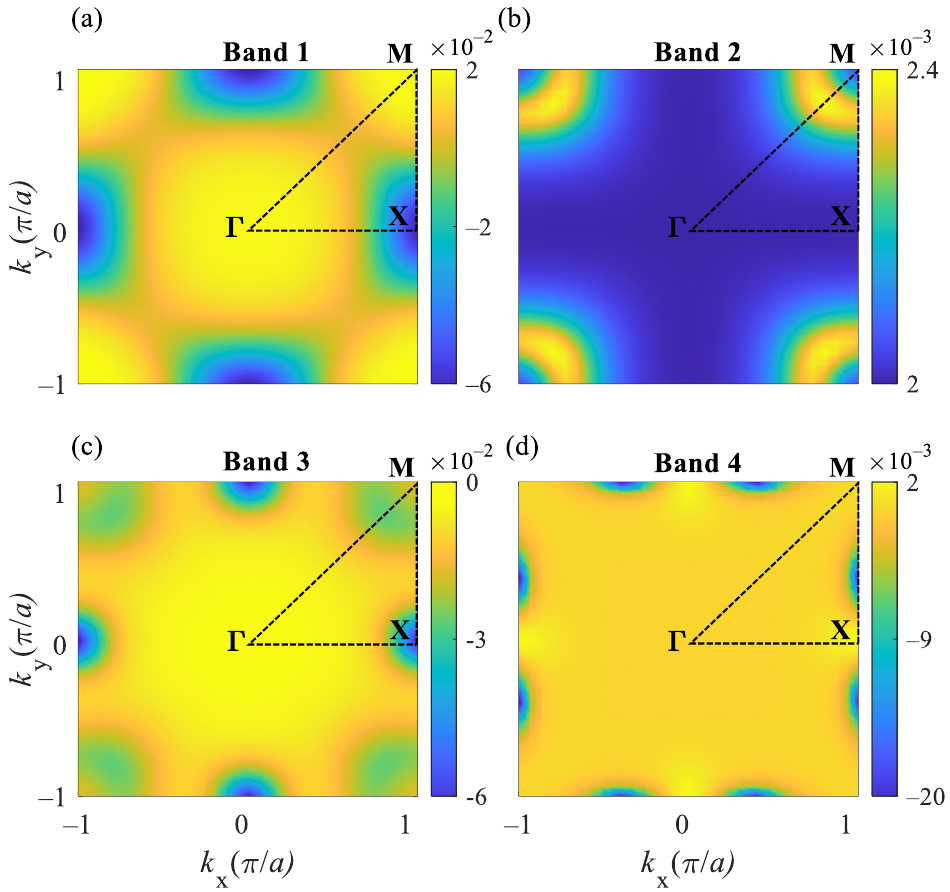}
\caption{\label{fig:S3} 
	Berry curvature distribution over $k_\textrm{x}$ and $k_\textrm{y}$ for the (a) first, (b) second, (c) third, and (d) forth bands of a YIG rod with radius $R_d^{0} = 0.11a$, $\epsilon_\textrm{n} = 15$, and $\mu = \bm{\mu}$. 
	The dashed triangles with the vortices of $\Gamma$, X, and M represent the irreducible BZ.}
\end{figure}

\begin{figure*}[t]
\includegraphics[width=1 \linewidth]{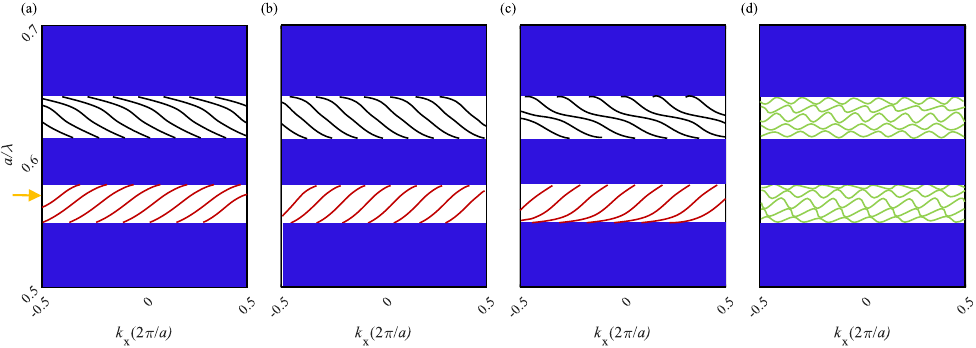}
\caption{\label{fig:S4} 
	Dispersion diagrams of the TW with 5 × 8 dielectric rods with $\epsilon_t = 16$ and 5 × 8 YIG rods. (a) $\eta_\textrm{P} = 40\%$ to the trivial lattice,
(b) $\eta_\textrm{P} = 40\%$ to the non-trivial lattice, (c) $\eta_\textrm{R} = 40\%$ to the trivial lattice, and (d) $\eta_\textrm{R} = 40\%$ to the non-trivial lattice.
	Black and red curves represent topological modes. Blue and green colors show bulk and defect modes, correspondingly. 
	The orange arrow shows $a/\lambda=0.57$.}
\end{figure*}

\begin{figure*}[t]
\includegraphics[width=1 \linewidth]{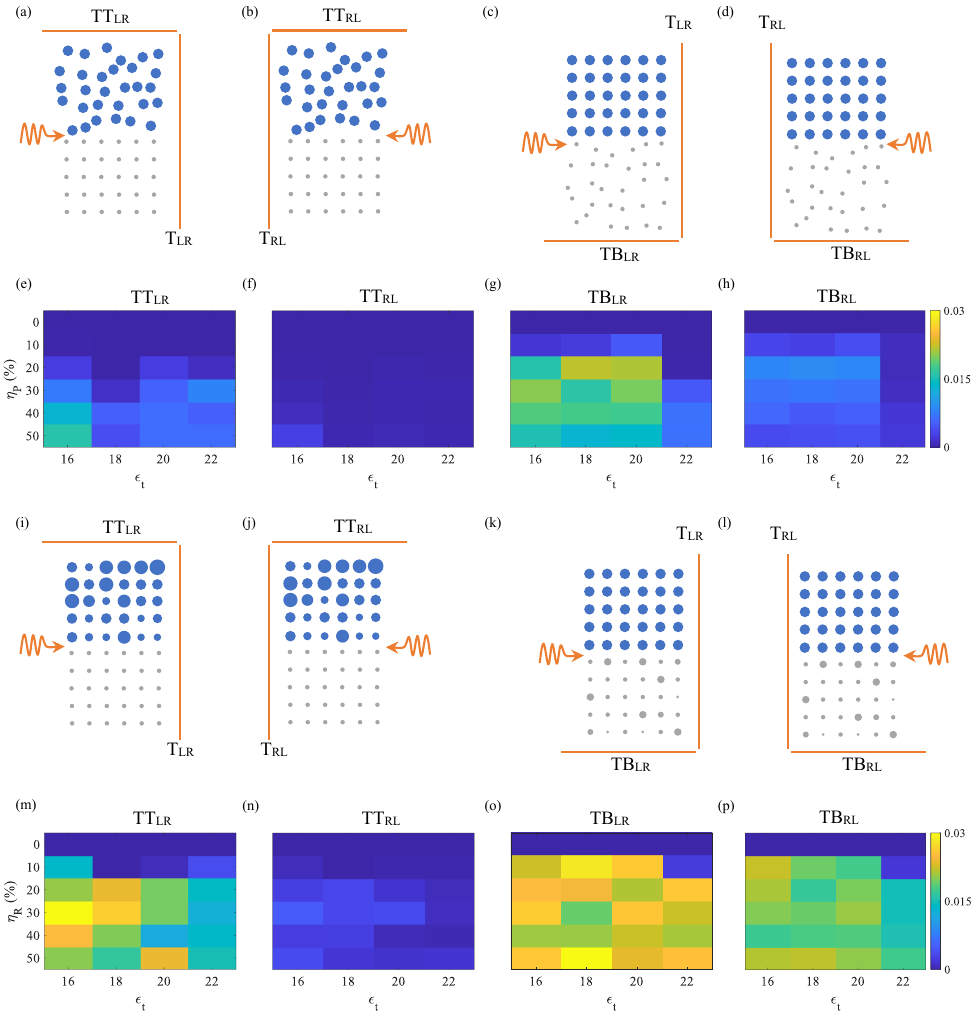}
\caption{\label{fig:S5} 
	The structure of the TW containing trivial lattice of  $5 \times 6$ dielectric rods in air and non-trivial lattice of the $5 \times 6$ YIG rods in air. 
	The TW under applying position disordering to the trivial lattice (a) and (b) and to the non-trivial lattice (c) and (d). 
	The TW under applying radius disordering to the trivial lattice (i) and (j) and to the non-trivial lattice (k) and (l). 
	For (a), (c), (i), and (k), Gaussian waves propagate from left to right and for (b), (d), (j), and (l), Gaussian waves propagate from right to left. 
	(e)--(h) are the simulation results for different $\epsilon_\textrm{t}$ for (a)--(d), respectively; (m)--(p) are the simulation results for different $\epsilon_\textrm{t}$ for (i)--(l), respectively.
	$a/\lambda = 0.570$ for $\epsilon_{\textrm{t}}=16, 18$, and $20$ and $a/\lambda = 0.550$ for $\epsilon_{\textrm{t}}=22$.}
\end{figure*}

The Berry curvature of the first four bands for the non-trivial lattice containing the YIG rod in air represents $C_4$ rotation symmetry over the first BZ; the $C_4$ rotation symmetry of the Berry curvatures is inherited from the same symmetry of the lattice (Fig.~\ref{fig:S3}). 
Also, due to the broken time-reversal symmetry and preserving inversion symmetry, the Berry curvatures reveal an even function over the wavevector, $F (-\textbf{k})$ = $F (\textbf{k})$. 
The integral of the Berry curvature for each band over the first BZ is proportional to the Chern number.
The Chern number for Band 1, Band 2, Band 3, and Band 4 are $0, +1, -2$, and $-1$, respectively.

\section{Dispersion diagram calculation}

The dispersion diagram is calculated by discretization of one of the master equations (Eqs.~(\ref{eq:13}) and (\ref{eq:14})) using the finite-element method (FEM) under applying Bloch boundary conditions (Eq.~(\ref{eq:15})) as follows

\begin{align}\label{eq:13}
	\frac{1}{\epsilon_\textrm{r}(\textbf{\textrm{r}})}\nabla \times \left[ \frac{1}{\mu_\textrm{r}(\textbf{\textrm{r}})} \nabla \times \textbf{\textrm{E}} \right] &= \left( \frac{\omega}{c} \right)^2\textbf{\textrm{E}}, \\
\label{eq:14}
	\frac{1}{\mu_\textrm{r}(\textbf{\textrm{r}})}\nabla \times \left[ \frac{1}{\epsilon_\textrm{r}(\textbf{\textrm{r}})} \nabla \times \textbf{\textrm{H}} \right] &= \left( \frac{\omega}{c} \right)^2\textbf{\textrm{H}},
\end{align}

\begin{equation}\label{eq:15}
\begin{split}
	\textbf{\textrm{E}}(\textbf{\textrm{r}}) & =\Phi_\textbf{\textrm{E}}(\textbf{\textrm{r}})e^{j\textbf{\textrm{K}}\cdot \textbf{\textrm{r}}}, \\
  \textbf{\textrm{H}}(\textbf{\textrm{r}})& =\Phi_\textbf{\textrm{H}}(\textbf{\textrm{r}})e^{j\textbf{\textrm{K}}\cdot \textbf{\textrm{r}}}, \\
\end{split}
\end{equation}

\noindent where \textbf{E}, \textbf{H}, \textbf{K}, $\epsilon_\textrm{r}, \mu_\textrm{r}, \omega, \Phi_\textbf{\textrm{E}}$, and $\Phi_\textbf{\textrm{H}}$ are the electric field, magnetic field, Bloch wavevector, relative permittivity, relative permeability, frequency, electric-periodic part of Bloch function and magnetic-periodic part of Bloch function, respectively.

The FEM works in real-space for calculating the dependent variables of \textbf{E}, \textbf{H}, $\omega$, and \textbf{K} by discretizing the master equation (either Eq.~(\ref{eq:13}) or (\ref{eq:14})). 
This method is useful for calculating photonic dispersion diagrams of disorder structures that are of experimental interest. 
Also, the method is used to calculate the dispersion diagram of arbitrary complex geometries with the material dispersion. 
In addition, discontinuities in the dielectric function do not affect the convergence of the method. 
The disadvantage of this method is requiring extensive computer memory.
The main part of the FEM is dividing the structure into small units which are called mesh elements. 
In two-dimensional FEM simulations, mesh elements are triangular based on the Delaunay algorithm. 
The FE method approximates independent variables using a linear combination of basis functions (shape functions) that are solved using an appropriate solver. 
Here, for calculating the dispersion diagrams of trivial and non-trivial lattices and the TW, FEM method embedded in COMSOL Multiphysics software is utilized.

To calculate the dispersion diagrams, a TW containing a supercell of $5 \times 8$ dielectric rods with $\epsilon_\textrm{t} = 16$ in air of a structure consisting of the $5 \times 8$ YIG rods in air is considered. 
The boundary conditions of supercells for $x_\textrm{min}$ and $x_\textrm{max}$ are Bloch boundary conditions and for $y_\textrm{min}$ and $y_\textrm{max}$ are continuity conditions. 
The dispersion diagrams of the TW under applying $\eta_\textrm{P} = 40\%$ to the trivial and non-trivial lattices  (Figs.~\ref{fig:S4}(a) and (b)) and $\eta_\textrm{R} = 40\%$ to the trivial lattice  (Fig.~\ref{fig:S4}(c)) represents two types of non-trivial mirrors. 
The non-trivial mirrors at $a/\lambda<0.6$ and $a/\lambda>0.6$ represent edge modes with positive and negative group velocities and C\textsubscript{g} of $+1$ and $-1$, respectively. 
The TW under applying $\eta_\textrm{R} = 40\%$ to the non-trivial lattice represents no unidirectional bandgap as shown in Fig.~\ref{fig:S4}(d), because defect modes occupying the bandgaps.

\section{Unidirectionality transmissions and scatterings}

We consider a TW containing $5 \times 6$ dielectric rods as a trivial lattice and $5 \times 6$ YIG rods as a non-trivial lattice. 
The FEM module of COMSOL Multiphysics software is utilized to numerically simulate the TW. 
Given disorderings to the trivial and non-trivial lattices, transmissions are recorded at the top (TT\textsubscript{LR} and TT\textsubscript{RL}) and bottom (TB\textsubscript{LR} and TB\textsubscript{RL}) of the TW, respectively. 
Under applying position disordering to the trivial lattice (Figs.~\ref{fig:S5} (a) and (b)), the scattering transmissions of TT\textsubscript{LR} and TT\textsubscript{RL} are approximately zero (Figs.~\ref{fig:S5}(e) and (f)). 
TB\textsubscript{LR} is higher than TB\textsubscript{RL} under applying position disordering to the non-trivial lattice that results in blocking the Gaussian wave that launched from the right side of the waveguide (Figs.~\ref{fig:S5}(c), (d), (g), and (h)).
Under applying radius disordering to the trivial lattice TT\textsubscript{LR} shows a higher value than TT\textsubscript{RL} (Figs.~\ref{fig:S5} (i), (j), (m), and (n)).
In contrast, applying radius disordering to the non-trivial lattice represents approximately the same values for TB\textsubscript{LR} and TB\textsubscript{RL}, revealing unidirectionality breaking (Figs.~\ref{fig:S5}(k), (l), (o), and (p)).

\section{Bott index}

The Bott index ($C_B$) is an integer number that can characterize both the ordered and disordered systems (with the broken time-reversal symmetry). 
The calculation procedure of the Bott index is as follows.
Here we propose a supercell of $6 \times 6$ YIG rods in air with anti-periodic boundary conditions in the $x$ and $y$ directions. 
The supercell is discretized into $M^2$ coordinated points. 
$N$ eigenmodes, $[E_{z}(x,y)]_{M^2 \times N}$, with eigenfrequencies below specific frequency $f$ are calculated using COMSOL Multiphysics software. 
Two Hermitian diagonal matrices of $[X]_{M^2 \times M^2}$ and $[Y]_{M^2 \times M^2}$ with $x$ and $y$ coordinates as diagonal elements are defined. 
Two diagonal matrices of $[U_{x}]_{M^2 \times M^2} = \textrm{exp}(i2\pi X/L_x)$ and $[U_{y}]_{M^2 \times M^2} = \textrm{exp}(i2\pi Y/L_y)$ are considered, where $L_x$ and $L_y$ are the lengths of the supercell in the $x$ and $y$ directions, respectively. 
Band-projected position matrices of $[\tilde{U}_{X}]_{N \times N} = [P^\prime]_{N \times M^2} [U_{x}]_{M^2 \times M^2} [P]_{M^2 \times N}$ and $[\tilde{U}_{Y}]_{N \times N} = [P^\prime]_{N \times M^2} [U_{y}]_{M^2 \times M^2} [P]_{M^2 \times N}$ are defined, where prims sign represents a transpose of the matrix.
$P$ is the projection matrix to all eigenmodes with frequencies below $f$. 
The square matrix is defined as $[A]_{N \times N} = [\tilde{U}_{Y}]_{N \times N} [\tilde{U}_{X}]_{N \times N} [\tilde{U}^\prime_{Y}]_{N \times N} [\tilde{U}^\prime_{X}]_{N \times N}$. 
A diagonal matrix $[B]_{N \times N}$ is introduced with diagonal elements containing the eigenvalues of $[A]_{N \times N}$. 
The Bott index at the frequency $f$ is introduced as

\begin{equation}\label{eq:16}
	C_{B}=\frac{1}{2\pi}\textrm{Im}\{\textrm{tr}(\textrm{ln}\{[B]_{N \times N}\})\},
\end{equation}

\noindent
where Im, tr, and ln are the imaginary part, trace, and Neperian logarithm, respectively.

\section{Electric field distribution}

We study the electric field distribution $|E_z(x,y)|$ of a structure consisting of the supercell containing $5 \times 10$ YIG rods in air at the bottom and a metal zone as a reflector at the top of it. 
The $|E_z(x,y)|$ distribution at each frequency and disordering is the avearge of 50 simulations. 
In Fig.~\ref{fig:S6}, $|E_z(x,y)|$ distributions show unidirectionality independent of the position disordering for the topological bandgaps of BG\textsubscript{n2} and BG\textsubscript{n3}; the directionality is broken at the other frequencies. 
$|E_z(x,y)|$ distributions for the structure under radius disordering demonstrate the unidirectionality of BG\textsubscript{n2} and BG\textsubscript{n3} at low radius disordering and unidirectionality breaking for the bandgaps at higher radius disordering, as presented in Fig.~\ref{fig:S7}. 
To some extent of increase in the radius disordering, the unidirectionality appears at the lower frequencies, revealing TAL. 
These results confirm the calculation results of Bott index in Fig.~4.

\begin{figure*}[t]
\includegraphics[width=1 \linewidth]{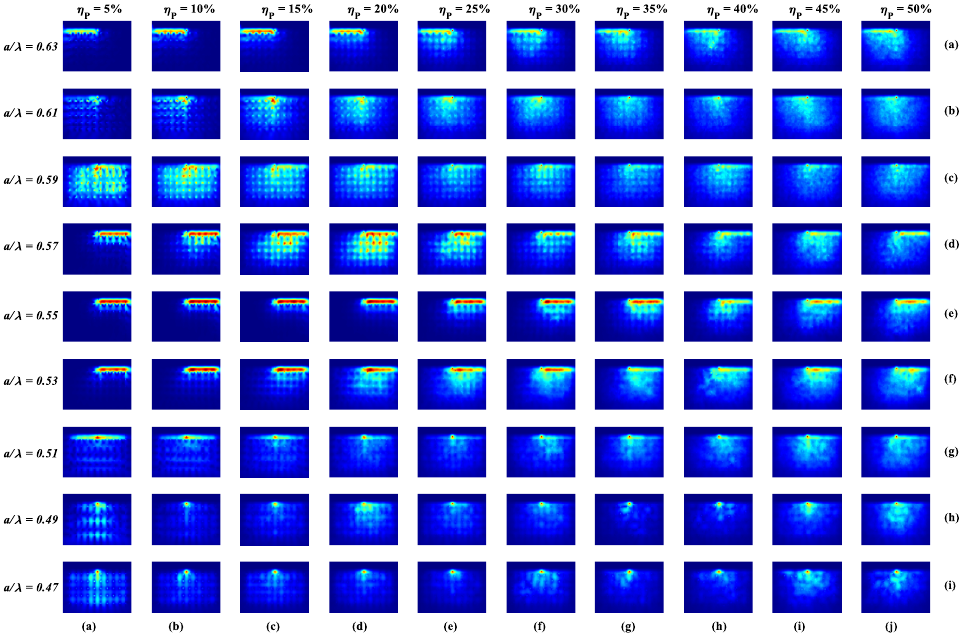}
	\caption{\label{fig:S6} The $|E_z(x,y)|$ distributions of a structure consists of the TW containing a non-trivial mirror of the $5 \times 10$ YIG rods in air and a metal bar as a trivial mirror at the top of it under position disordering at different frequencies.}
\end{figure*}

\begin{figure*}[t]
\includegraphics[width=1 \linewidth]{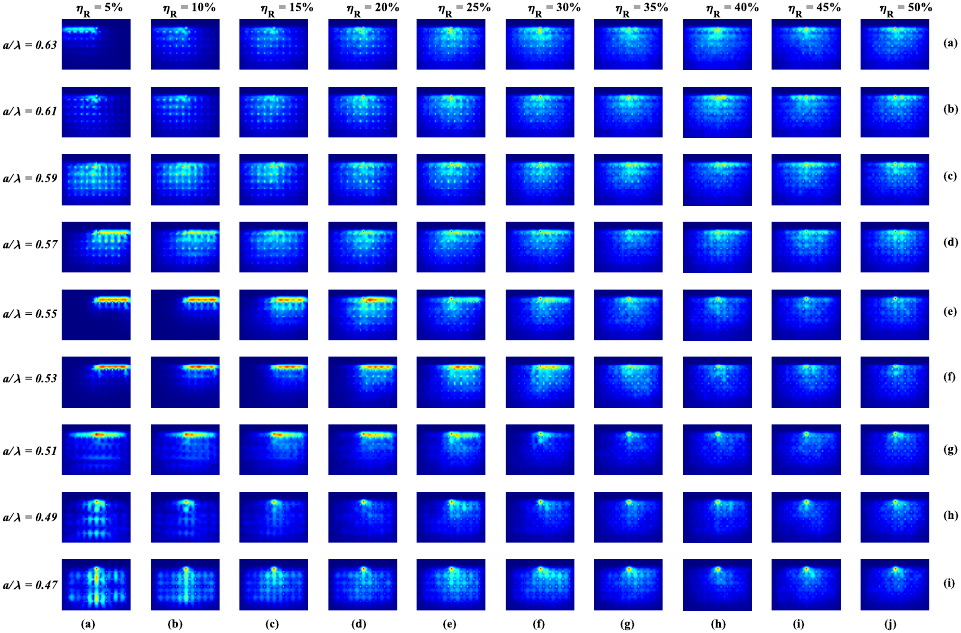}
	\caption{\label{fig:S7} The $|E_z(x,y)|$ distributions of a structure consists of the TW containing a non-trivial mirror of the $5 \times 10$ YIG rods in air and a metal bar as a trivial mirror at the top of it under radius disordering at different frequencies.}
\end{figure*}

\section{Eigenfrequency analysis}

This section describes the eigenfrequency analysis of the non-trivial lattice. 
The lattice consists of $6 \times 6$ YIG rods in air. 
The eigenfrequency of the lattice under applying the radius disorder to the YIG rods as a function of eigenfrequency is computed using the FE method embedded in COMSOL Multiphysics software. 
The eigenfrequencies are computed for the average of an ensemble of 50 simulations. 
The boundary conditions are anti-periodic boundary conditions in both the $x$ and $y$ directions. 
The computed eigenfrequencies illustrate three main bandgaps of BG\textsubscript{n1}, BG\textsubscript{n2}, and BG\textsubscript{n3} are in gray with no disorder as illustrated in Fig. \ref{fig:S8}. 
By increasing the radius disorder, the eigenfrequencies from the upper edge of the bandgaps of BG\textsubscript{n1} and BG\textsubscript{n2} fall down to the bandgaps and experience red shifts. 
BG\textsubscript{n1} and BG\textsubscript{n2} disappear at $\eta_\textrm{R} = 50\%$ and $\eta_\textrm{R} = 20\%$. 
By increasing the radius disorder, eigenfrequencies experience red and blue shifts for BG\textsubscript{n3}. Also, BG\textsubscript{n3} disappears at $\eta_\textrm{R} = 10\%$.

\begin{figure*}[t]
\includegraphics[width=1 \linewidth]{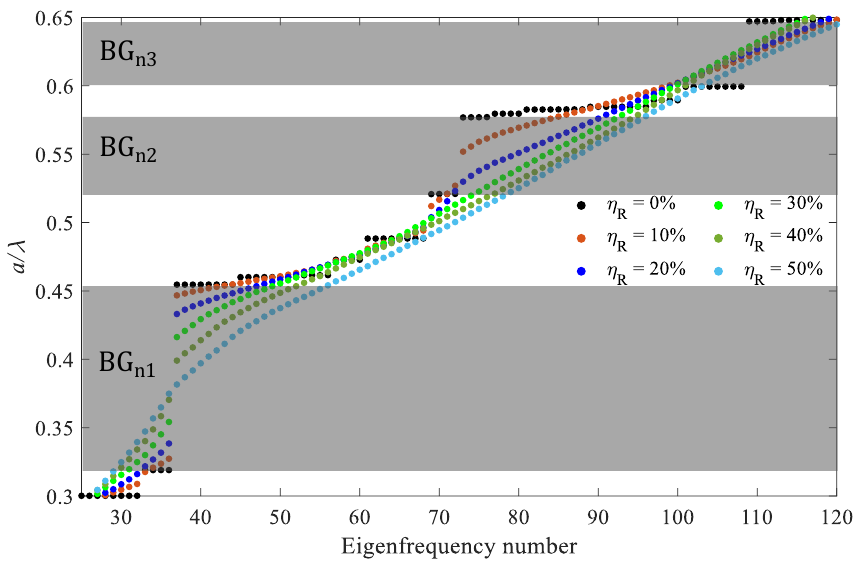}
\caption{\label{fig:S8} The eigenfrequencies of a non-trivial lattice containing of the $6 \times 6$ YIG rods in air under applying the radius disorder. 
Gray regions show the bandgaps of BG\textsubscript{n1}, BG\textsubscript{n2}, and BG\textsubscript{n3}.}
\end{figure*}

\end{document}